\documentclass[prb,twocolumn,floatfix]{revtex4-1}

\usepackage{color}
\usepackage{graphicx}
\usepackage{amsmath}
\newcommand{\be}{\begin{equation}}
\newcommand{\ee}{\end{equation}}

\begin{document}

\title{Effect of shear-coupled grain boundary motion on coherent precipitation}


\author{M. Weikamp}
\affiliation{Institute for Energy and Climate Research IEK-2, Forschungszentrum J\"ulich GmbH, 52425 J\"ulich, Germany}
\author{R. Spatschek}
\affiliation{Institute for Energy and Climate Research IEK-2, Forschungszentrum J\"ulich GmbH, 52425 J\"ulich, Germany}

\begin{abstract}
We examine the interaction between precipitates and grain boundaries, which undergo shear-coupled motion.
The elastic problem, emerging from grain boundary perturbations and an elastic mismatch strain induced by the precipitates, is analysed.
The resulting free elastic energy contains interaction terms, which are derived numerically via the integration of the elastic energy density. 
The interaction of the shear-coupled grain boundary and the coherent precipitates leads to potential elastic energy reductions. 
Such a decrease of the elastic energy has implications on the grain boundary shape and also on the solubility limit near the grain boundary.
By energy minimisation we are able to derive the grain boundary shape change analytically. 
We apply the results to the Fe-C system to predict the solubility limit change of cementite near an $\alpha$-iron grain boundary.
\end{abstract}

\date{\today}

\maketitle
\section{Introduction}
The understanding and the associated targeted influencing of mechanical properties of steels and alloys is an important and necessary part of materials science.
The process of precipitation and consequently the presence of secondary phases with different properties is an important part of a microstructure.
Cementite, for example, is very brittle but also hard and can reduce the effectiveness of an alloy or steel as it can act as crack initiator \cite{Park:1979aa}.
Precipitates in general also influence dislocation movement and can therefore also strengthen the material (precipitation hardening \cite{Ardell:1985aa}).
In an earlier study \cite{Spatschek:2016aa} the precipitation of hydrides near surfaces has been investigated, showing that elastic relaxation near free surfaces leads to significant differences concerning the phase stability compared to bulk precipitation.
The interaction of precipitates and a free surface leads to a reduction of the elastic energy and therefore results in a change of the solubility limit.
A transfer to interfaces has been shown in Ref.~\onlinecite{Weikamp:2018aa}, where a grain boundary is described as a mesoscopic layer with different elastic properties compared to the bulk.
In this representation the grain boundary acts as a generally non-free surface and an influence on the solubility limit is observable. 
However, such an effective picture does not consider microscopic details of strengthening or stress release mechanisms.
The aim of the present article is therefore a more explicit consideration of stress release mechanisms due to morphological rearrangements of grain boundaries, in particular through shear-coupled motion, in conjunction with precipitate formation.

Shear-coupled motion of grain boundaries describes the normal motion of a grain boundary while the grains are sheared parallel to each other.
This reversible interaction has been known for quite some time, first theoretically predicted by Read and Shockley in 1950, see Ref.~\onlinecite{Read:1950aa}.
The authors derived, that for low angle symmetric tilt grain boundaries the collective movement of edge dislocations leads to the normal grain boundary motion as a response to shear stress.
First experimental evidence was found a few years laters in zinc bi-crystals \cite{Li:1953aa,Bainbridge:1954aa}.
The theoretical work of Cahn et al.\cite{Cahn:2004aa} in 2004 lead to a unified approach to describe the mechanism of pure sliding and shear-coupled motion.
Molecular dynamics (MD) simulations on symmetrical [001] tilt boundaries revealed\cite{Cahn:2006aa,Cahn:2006ab}, that the underlying phenomenon is also applicable to high angle grain boundaries, which cannot be considered as array of isolated dislocations. 
In Ref.~\onlinecite{Adland:2013aa} the shear-coupling behaviour and misorientation angle dependence has been investigated by phase field crystal simulations, showing also a transition from coupled motion to sliding at higher homologous temperatures.
Further experimental \cite{Molodov:2011aa} and numerical\cite{Trautt:2012aa} studies show, that shear-coupled motion also occurs for more realistic and complex asymmetric grain boundaries.
Another investigation\cite{Gorkaya:2011aa} reveals, that mixed-mode grain boundaries with a twist component also undergo shear-coupled motion correlated to the tilt-fraction of the grain boundary.
In Ref.~\onlinecite{Cheng:2016aa} the $\Sigma 5 (310)$ grain boundary in Al has been investigated by MD simulations.
The interaction between shear and normal grain boundary motion has been observed, where the multiplicity of the grain boundary leads to different grain boundary structures after thermal relaxation.
Another MD study \cite{Niu:2016aa} shines light on the different modes of dislocation movement of a [001] grain boundary in bcc W, coming to the conclusion that the $\langle110\rangle$ mode of the dislocation movement leads to easier shuffling of the atoms at the grain boundary.
Further publications \cite{Geslin:2015aa,Xu:2016aa} have investigated the interaction between shear-coupled grain boundary motion and a lamellar precipitate, which engulfs the boundary, as such an arrangement is energetically favorable.
A linear stability analysis and phase field crystal simulations show, that grain boundaries can become unstable and break-ups occur.
In Ref.~\onlinecite{Xu:2016aa} the authors show similar simulation results also for a spherical inclusion. 
All of these studies show, that any complex grain boundary structure can exhibit shear-coupled movements, and the later works indicate an influence on precipitates.

Based on these works, the present article aims to establish a quantitative link between shear coupled grain boundary relaxation and (coherent) precipitation from a thermomechanical perspective.
It turns out that this combination can locally alter the thermodynamic landscape and therefore favour precipitation near grain boundaries, in agreement with the observations mentioned above.
To concisely demonstrate the concept of the interaction between shear-coupled grain boundaries and precipitates, we 
follow the perturbation analysis published in Ref.~\onlinecite{Karma:2012aa}.
Spherical precipitates are introduced in the vicinity of the grain boundary and the correlation between the elastic fields is investigated.
The interaction between precipitates and grain boundary leads to an elastic interaction term, which allows the system to lower its elastic free energy.
It can therefore be favourable for precipitates to be located at specific locations near a shear-coupled grain boundary.
The consequence is a solubility limit change at these positions.

This article is structured as follows.
After a brief discussion of shear coupled motion in Section \ref{sec2}, we derive the elastic energy of an independent shear-coupled grain boundary and isolated precipitates in Section \ref{sec3}. 
In Sec.~\ref{sec:Interaction} the interaction is considered, leading to new correlation terms.
The interaction terms potentially lower the free elastic energy of the system.
This induces a grain boundary shape change as a result of the energy minimisation, as discussed in Section \ref{sec4}.
Moreover this leads to a  solubility limit change and therefore a local modification of phase diagrams for alloys, which is discussed in Sec.~\ref{sec:solubility}.

\section{Shear-coupled motion}
\label{sec2}

\begin{figure}
\begin{center}
\includegraphics[width=8.6cm]{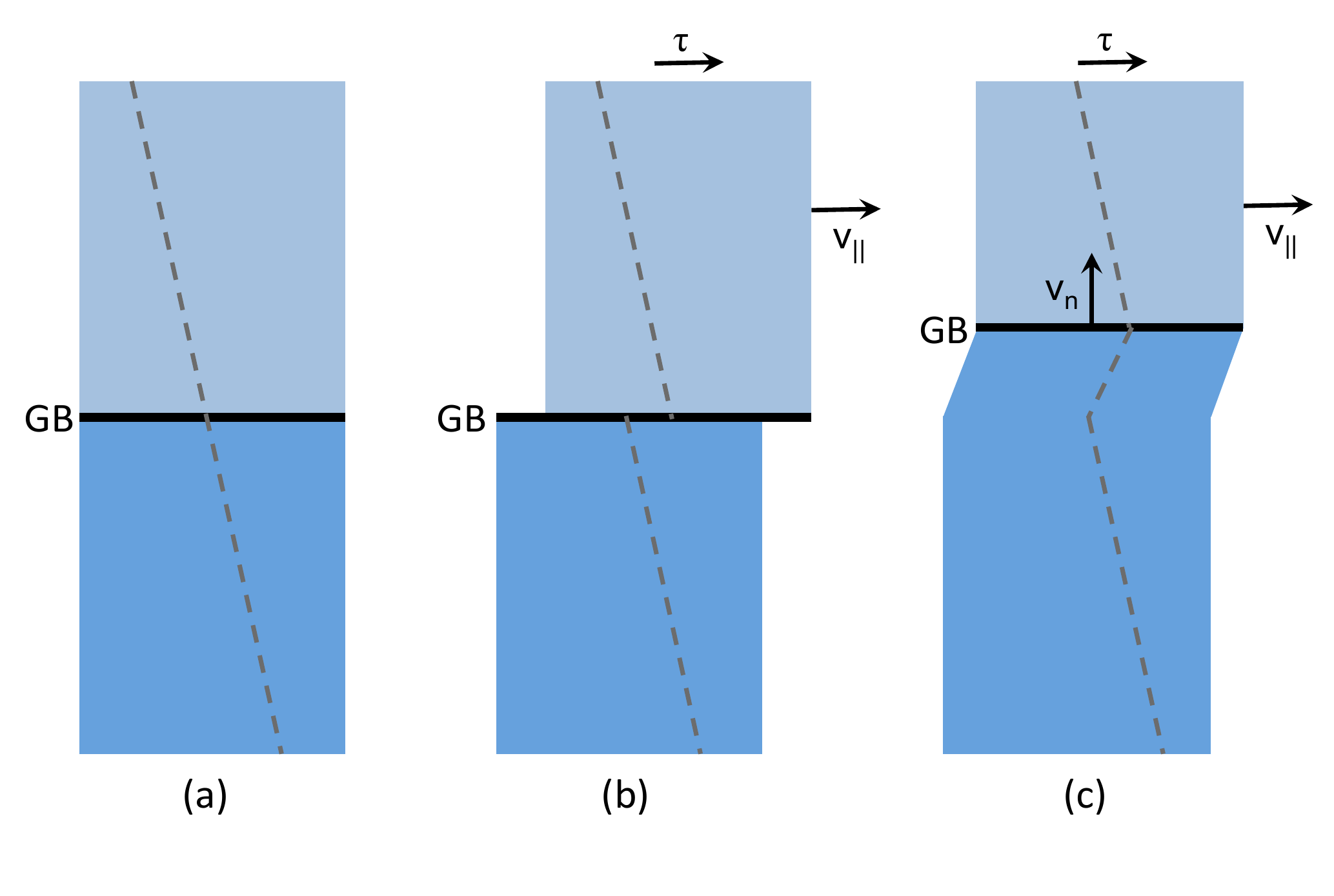}
\caption{
Illustration of sliding and shear coupled grain boundary motion, for a horizontal grain boundary (GB) as shown in the reference state (a).
The upper grain is sheared by a stress $\tau$, which leads to a tangential velocity $v_{||}$. For pure sliding (b), the grain boundary maintains its position, whereas for shear-coupling (c), grain boundary normal motion with velocity $v_n$ occurs.
}
\label{fig:sketch_SCM}
\end{center}
\end{figure}

A general law of grain boundary motion caused by shear stress is given in Ref.~\onlinecite{Cahn:2006aa}, describing the tangential grain velocity $v_{||}$ as
\be \label{eq:slidingandshearcoupling}
v_{||} = S \tau + \beta v_n.
\ee
The tangential sliding velocity is thus a resulting combination of sliding and normal grain boundary motion $v_{n}$. 
The first term on the right side describes the sliding motion of a grain due to a shear stress $\tau$ acting at the top grain via a sliding coefficient $S$.
The second term captures the coupling to the normal grain boundary velocity $v_n$ via the coupling factor $\beta$.
A sketch to illustrate the two mechanisms is shown in Fig.~\ref{fig:sketch_SCM}.
In this work we focus on the case of pure shear-coupled motion, therefore using the relation
\be \label{eq:shearcoupling}
v_{||} = \beta v_n.
\ee 
For symmetric [001] tilt grain boundaries, the coupling factor is only dependent on the crystallographic landscape, if the temperature is well below the melting temperature, $T < 0.7 \ T_M$, as shown in Ref.~\onlinecite{Cahn:2006aa}.
This grain boundary behaviour is applicable for low and high angle tilt boundaries\cite{Cahn:2004aa,Cahn:2006aa,Cahn:2006ab}.
For low misorientation angles near zero ($\theta \rightarrow 0$) one uses
\be \label{eq:shearcoupling_factor1}
\beta_{\langle100\rangle} = 2 \tan \left( \frac{\theta}{2} \right).
\ee
This relation changes to a second branch for misorientation angles approaching the opposite limit, $\theta \rightarrow 90^\circ$, which leads to the relationship
\be \label{eq:shearcoupling_factor2}
\beta_{\langle110\rangle} = -2 \tan \left(\frac{\pi}{4} - \frac{\theta}{2} \right).
\ee
These two relations originate from two different slip directions of the grain boundary dislocations and can be derived using the Frank-Bilby equation\cite{Cahn:2006ab}.
The transition angle at which the coupling factor changes from the $\langle100\rangle$ mode to the $\langle110\rangle$ mode is dependent on the temperature, as reported in Ref.~\onlinecite{Cahn:2006aa} for Copper.
For aluminum, the coupling factor is reported to be independent on temperature and remains in the $\langle100\rangle$ mode at a high misorientation angle, according to MD simulations\cite{Ivanov:2008aa}.   
The coupling factor $\beta$ has been analysed and confirmed multiple times via experiments \cite{Gorkaya:2009aa, Molodov:2007aa} and simulations \cite{Homer:2013aa,Huter:2015aa} for different materials and symmetric tilt boundaries.

\section{Elastic energy}
\label{sec3}

\begin{figure}[]
\begin{center}
\includegraphics[width=8.6cm]{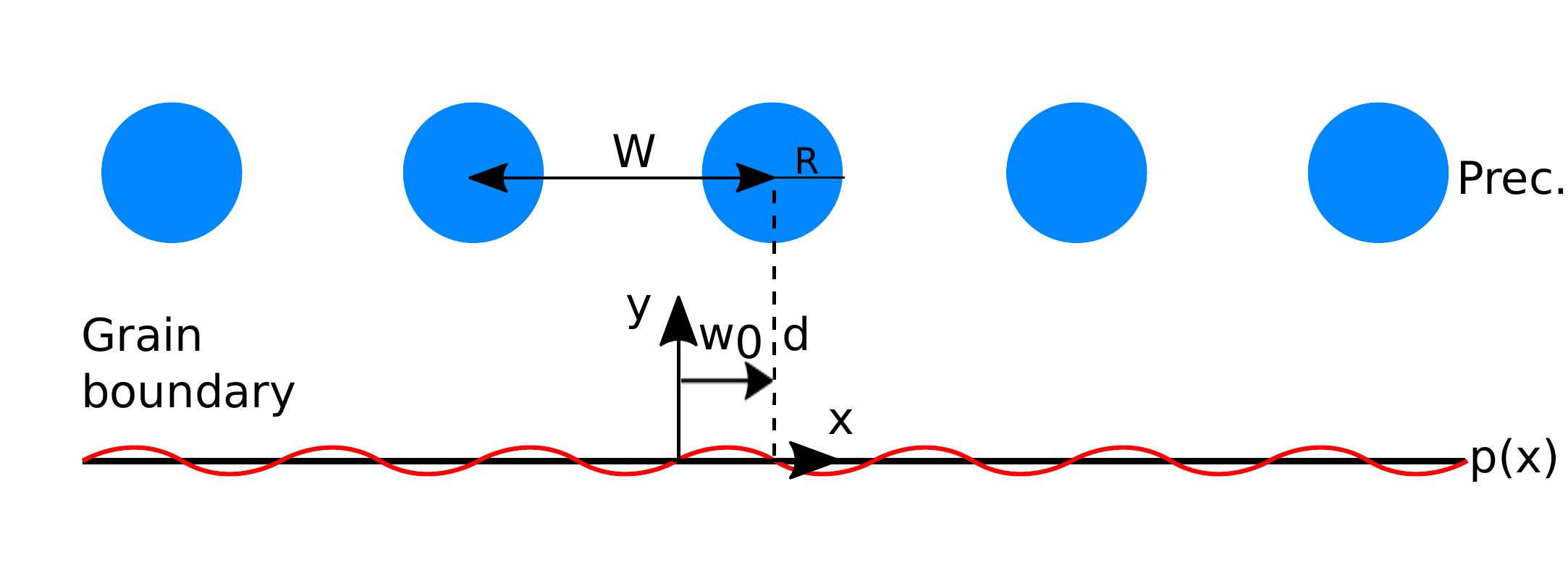}
\caption{Illustration of a (single sine mode) perturbation of the grain boundary via the function $p(x)$.
Additionally, precipitates in an array with interval length $W$ are located in a distance $d$ on top of the grain boundary, measured by the line between $y=0$ and the centre of the precipitates.
A horizontal shift of the precipitate array is defined by the parameter $w_0$, thus the center of one precipitate is located at $x=w_0$.}
\label{fig:sketch_setup}
\end{center}
\end{figure}

For simplicity, we investigate a two-dimensional setup, which contains a grain boundary and circular (cylindrical in three dimensions) precipitates, see Fig.~\ref{fig:sketch_setup} for a sketch.
The grain boundary is allowed to undergo shear-coupled motion, while the coherent precipitates of radius $R$ are located in the surrounding matrix above or below the grain boundary.
The precipitates are assumed to have an isotropic elastic lattice mismatch (eigenstrain $\varepsilon_0$) with the matrix phase, leading to the appearance of coherency stresses. 
A morphological perturbation of the grain boundary due to shear coupled motion leads to an increase of the elastic energy and to an interaction between the boundary and the precipitates.
In order to analyse the described problem, we follow the approach of Karma et al. \cite{Karma:2012aa}.
We use linear isotropic elasticity where the elastic constants of the matrix-phase and the precipitates are assumed to be equal.
The approach involves the derivation of displacement fields, the calculation of the free elastic energy density and finally an integration to yield the free elastic energy.
The final result of the free elastic energy of the depicted setup will consist of three main parts, the elastic grain boundary energy, the elastic energy of the precipitates and the interaction energy between grain boundary and inclusions.
%
\subsection{Grain boundary energy}
An initially straight grain boundary, expressed by the function $y=0$, is deformed by
\be \label{eq:perturbation}
p(x) = \sum_{k=0}^\infty \left[ a\left(k \right) \cos\left(k x \right) + b\left(k\right) \sin\left(k x\right) \right],
\ee
where $a(k)$ and $b(k)$ are wavenumber dependent Fourier series amplitudes of the perturbation.
In order to utilise this form of the perturbation, a periodic system has to be considered.
Although we focus on an isolated grain boundary first, the periodicity constraint has the implication, that the precipitates are arranged periodically as well.
Their vertical positions are defined by the parameter $d$, the distance between the unperturbed grain boundary and the center of the precipitates, and $W$ is the lateral spacing between them.  
This parameter defines the periodicity of our setup and therefore leads to a direct definition of the wavenumbers used in the Fourier series,
\be \label{eq:wavenumber}
k = \frac{2 \pi m}{W},\quad m =0,1,2,... \ .
\ee 
The last free parameter is the shift of the precipitates in horizontal direction, defined by $w_0$ (after energy minimization this parameter will drop out, reflecting the translational invariance of the problem).
For $w_0 = 0$, one precipitate is located symmetrically above $x=0$.
The deformation of the grain boundary via the function $p(x)$ in the context of shear-coupling is not only a movement in the normal direction but also implies a tangential displacement of the grains.
The central boundary condition of the elastic problem reflects the shear-coupled motion relation, Eq.~(\ref{eq:shearcoupling}), by expressing the lateral displacement via \cite{Karma:2012aa}
\begin{equation} \label{eq:perturbationfunction}
u_x^+(x,0) - u_x^-(x,0) = \beta p(x)
\end{equation}
up to linear order in $p(x)$, which serves as expansion parameter.
Here, one has to distinguish between the displacement components in the upper ($u^+_i, \ y>0$) and lower ($u^-_i, \ y<0$) domain.
%
%
The energy density is given by
\be \label{eq:ElasticEnergyDensity}
f = \frac{1}{2} \lambda \varepsilon_{kk}^2 + \mu \varepsilon_{ij}^2
\ee
in terms of the strain tensor $\varepsilon_{ij}=(\partial_j u_i + \partial_i u_j)/2$, using the shear modulus $\mu$ and the Lam\'e coefficient $\lambda=2\mu\nu/(1-2\nu)$ with the Poisson ratio $\nu$.
The elastic free energy is obtained by integration of the elastic energy density.
In horizontal direction the integration is determined by the periodic length unit $L = N \cdot W$, 
\be
F^\mathrm{GB} = \int_0^L dx \int_{-\infty}^{\infty} dy \ f(u_x,u_y) ,
\ee
with $N$ being the number of precipitates.
Details of the integration are shown in the supplemental material \cite{SMF}; the final result for an isolated grain boundary without precipitates reads
\be \label{eq:GBenergy}
F^\mathrm{GB} = \sum_{k=0}^{\infty} \left( \frac{\mu W N}{8(1-\nu)} \beta^2 k a^2(k) + \frac{\mu W N}{8(1-\nu)} \beta^2 k b^2(k) \right),
\ee
which corresponds to the result presented in Ref.~\onlinecite{Karma:2012aa} for a single cosine mode.

\subsection{Energy of precipitates}
We assume that the precipitates have a purely dilatational or compressive isotropic eigenstrain $\varepsilon_0$ with respect to the mother phase (hence the equilibrium strain in a stress free precipitate phase would be $\varepsilon_{ij} = \varepsilon_0 \delta_{ij}$).

The bulk free elastic energy of the two-phase system with coherent interface between matrix and precipitate of radius $R$ is according to the Bitter-Crum theorem \cite{Fratzl:1999aa}
\be \label{eq:SinglePrecEnergy}
F^\mathrm{prec} = \pi R^2 \varepsilon_0^2 \frac{E}{(1-\nu)},
\ee
with $E$ being the Young's modulus, which is related to the previous elastic parameters via $E = 2 \mu (1+\nu)$.
The elastic energy depends only on the total volume/area of the precipitate and not on the geometric arrangement.
For the considered case of vanishing elastic constant contrast between the phases and isotropic elasticity and eigenstrain, multiple precipitates do not interact in the bulk.
Therefore, $N$ of them lead to an increase of elastic energy by the factor $N$ (provided that they do not overlap, $R<W/2$),
\be \label{eq:MultiplePrecEnergy}
F^\mathrm{prec} = N \pi R^2 \varepsilon_0^2 \frac{E}{(1-\nu)}.
\ee

\subsection{Interaction and total elastic energy} 
\label{sec:Interaction}
The grain boundary modes and the precipitates have been considered separately up to this point.
Due to linearity, the total displacement, strain and stress fields are the sum of the contributions from the grain boundary and the precipitates.
Since energy is quadratic in strain, a cross term between the two contribution emerges, additionally to the self energies, which have been determined in the preceding sections.


The integration of the cross term energy density has been performed by a numerical integration method. 
By changing the modelling parameters on multiple scales, a reliable closed  expression for the interaction energy has been determined, which reads\cite{SMF}
\begin{eqnarray}
F^\mathrm{int} &&=  \frac{\Pi}{2}  \frac{E}{1-\nu}   \varepsilon_0  \beta R^2   N \sum_{k=0}^{\infty} \exp(- k  d)  k \Big[   a(k)    \sin(k  w_0)  \nonumber \\
&&-  b(k)  \cos(k  w_0)   \Big]. \label{eq:mixenergy}
\end{eqnarray}
This expression is valid for precipitates which are located in the upper grain, $y>0$, and which do not intersect with the grain boundary, $d>R$. 
Similarly, precipitates in the lower grain lead to the same expression with opposite sign.
The parameter $\Pi$ in Eq.~(\ref{eq:mixenergy}) is a constant which is approximately $\pi$ but deliberately left uncertain due to potential minor numerical inaccuracies. 
In the following we will assume $\Pi = \pi$.


The total elastic free energy of the system with infinite amount of grain boundary perturbations and $N$ precipitates is given as the sum of all contributions,
\be  \label{eq:TotalEnergy_2}
F = F^\mathrm{prec} + F^\mathrm{GB} + F^\mathrm{int}.
\ee

\section{Interpretation}
\label{sec4}

\subsection{Energy minimisation}

Inspection of the interaction energy (\ref{eq:mixenergy}) shows that the energy can either be increased or decreased, leading to a repulsive or attractive interaction for a fixed grain boundary shape $p(x)$.
This becomes obvious by the fact that both the eigenstrain $\varepsilon_0$ and the shear coupling factor $\beta$ can be either positive or negative.
For the first, it depends on the relative volume change of the precipitate in comparison to the matrix, for the latter it follows directly from Eqs.~(\ref{eq:shearcoupling_factor1}) and (\ref{eq:shearcoupling_factor2}).

The main novel aspect emerges from the fact that if shear coupled rearrangements of the grain boundary are possible, it can arrange such that the total free energy is minimized.
For illustrational purposes we follow this under the assumption that the energy contributions in Eq.~(\ref{eq:TotalEnergy_2}) are dominant, hence we suppress higher order corrections of the grain boundary energy $F^\mathrm{GB}$ from the perspective of the perturbative approach with small shape deviations $p(x)$, and also assume that a bare grain boundary energy, which depends on the grain boundary length, is subdominant to the elastic energy contributions which arises from the shear coupling.
Then, energy minimization with respect to the Fourier amplitudes $a(k), b(k)$ gives\cite{SMF}
\be \label{eq:totalenergy_minimised_simplified}
F^\mathrm{min} = F^\mathrm{prec} \times
\left( 1 - \frac{\pi^2 (1+\nu)}{2} \frac{R^2}{W^2  \sinh^2\left(\frac{2 \pi d}{W} \right)}  \right).
\ee
Obviously, this energy is lower than the precipitates energy near planar grain boundaries ($F^{GB}=0$), hence a short-ranged attractive interaction with an exponential asymptotic decay emerges.
This implies that the precipitate formation should occur more likely in the vicinity of the grain boundary.

An interesting outcome of this energy description is, that the shear coupling factor $\beta$ drops out.
The misorientation angle of the grain boundary no longer influences the energy, and therefore the behavior is expected to be generic for a wide range of grain boundaries. 
Due to the energy minimization the lateral shift $w_0$ cancels, which means that the ``phase'' of the perturbations aligns properly to the location of the precipitates, as will be discussed in more detail in the following section.
As the eigenstrain $\varepsilon_0$ appears quadratically in the energy expression, the attraction of the precipitates to the grain boundary is independent of the sign of the lattice mismatch.

\begin{figure}
\begin{center}
\includegraphics[trim={0 0 0 0}, width=8.6cm]{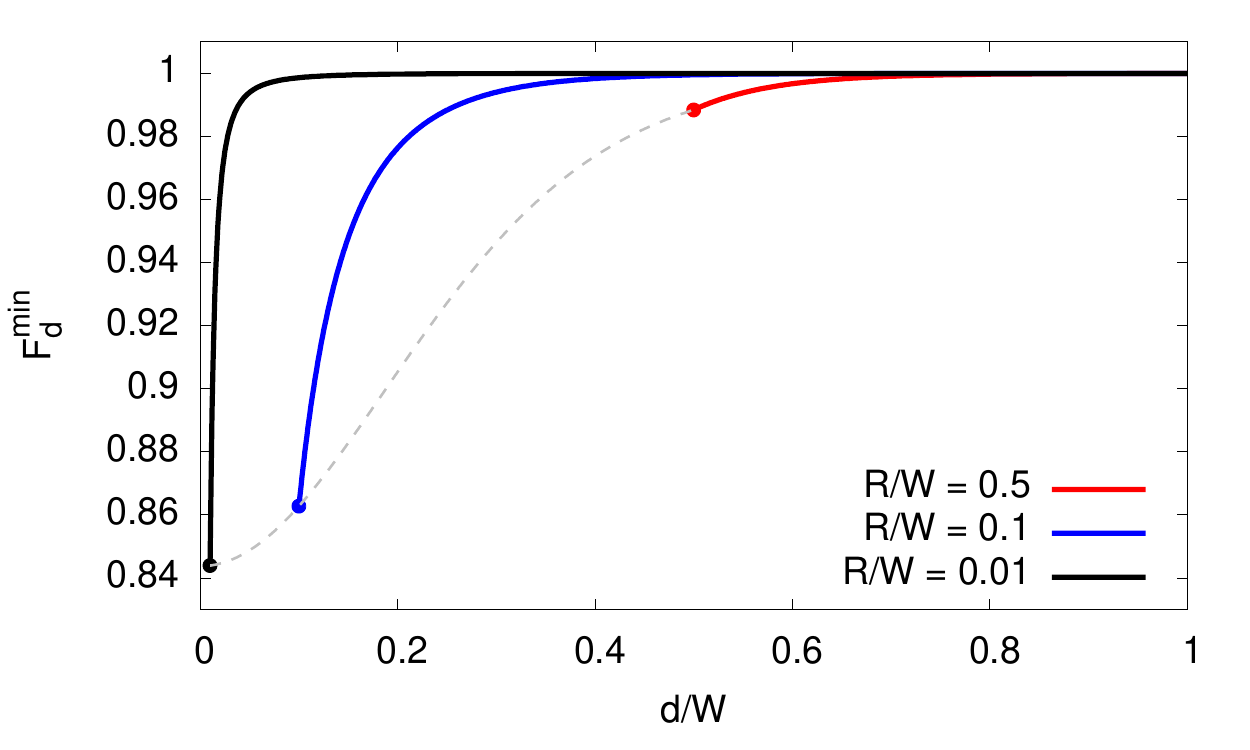}
\caption{Dimensionless free elastic energy $F^\mathrm{min}_d = F^\mathrm{min} / F^\mathrm{prec}$ as function of the (dimensionless) precipitate distance to the grain boundary, $d/W$ for $\nu=1/4$.
Different combinations of $R/W$ are chosen to illustrate the energy reduction dependent on the precipitate radius $R$. 
The curves are starting at positions indicated by a dot, which reflects the condition $d \geq R$, such that the precipitates do not intersect with the grain boundary. 
The limit of this condition, $d=R$, is shown by the dashed curve. }
\label{fig:Energy_plot}
\end{center}
\end{figure}

In Fig.~\ref{fig:Energy_plot}, the dimensionless free elastic energy $F^\mathrm{min}_d = F^\mathrm{min} / F^\mathrm{prec}$ dependent on $d/W$ (ratio grain boundary-precipitate to inter-precipitate distance) is shown.
%
It expresses the reduction of elastic energy when the precipitates approach the shear-coupled grain boundary.
Contrary, when $F_d^\mathrm{min}$ becomes unity for large separations, the interaction of inclusions and interface is negligible. 
Different ratios of $R/W$ are used to illustrate the scaling of the free elastic energy with the precipitates radius.
The curves are starting at positions indicated by a dot, marking the condition $d \geq R$, as otherwise the precipitates would intersect with the interface.
We have confirmed numerically that such an intersection is energetically unfavorable.
It is immediately visible, that the precipitates favour small distances to the grain boundary (attractive interaction).
On the other hand, when the precipitate radius $R$ increases, the curves are shifted to the right and the minimum value of $F^\mathrm{min}_d$ becomes larger.
The system therefore favours small precipitates, as they can be closer to the grain boundary.
Similarly, an increasing horizontal distance $W$ between the precipitates is favourable, which expresses an effective mutual repulsion of the precipitates near the grain boundary. 

\subsection{Change of the grain boundary shape}

\begin{figure}[]
\begin{center}
\includegraphics[width=8.6cm]{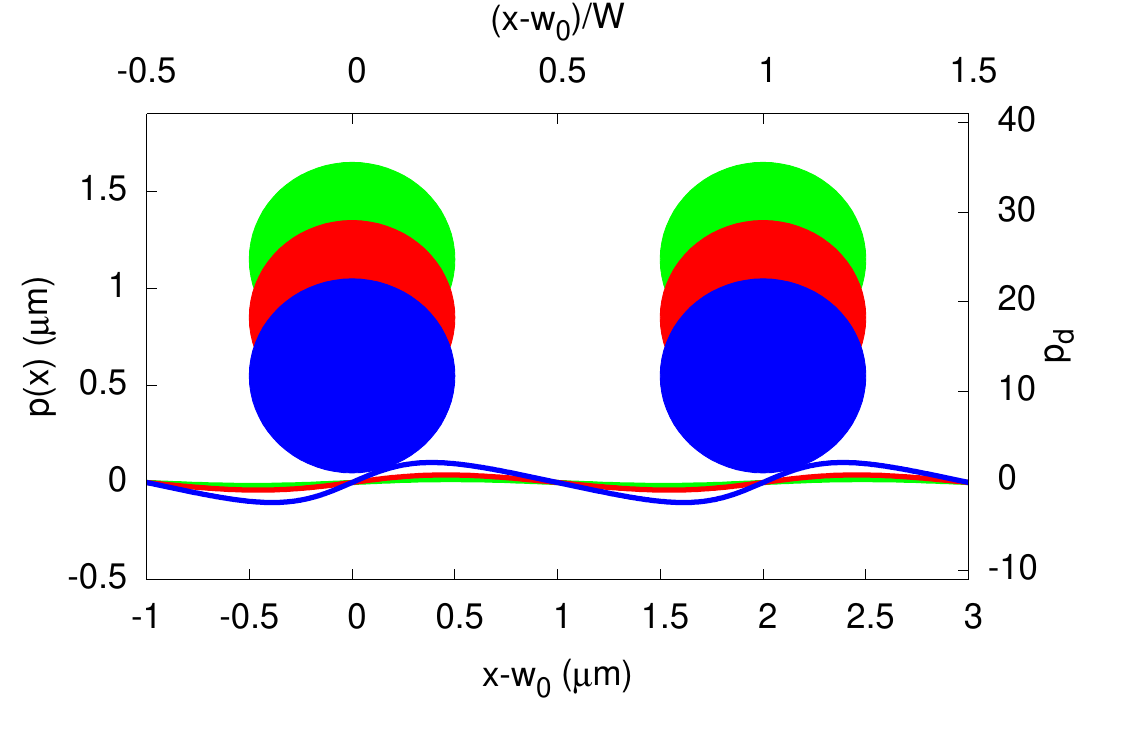}
\caption{Grain boundary shapes $p(x)$ for three different distances between the precipitates and the grain boundary, namely $d_{blue} = 5.5 \cdot 10^{-7}$ m, $d_{red} = 8.5 \cdot 10^{-7}$ m and $d_{green}= 1.15 \cdot 10^{-6}$ m.
The used parameters are $E = 175$ GPa, $\nu$ = 0.25, $R = 5 \cdot 10^{-7}$ m, $W = 2 \cdot 10^{-6}$ m, $\beta = 0.07$ 
and $\varepsilon_0 = 0.02$. 
Corresponding dimensionless values are shown on the secondary axis.
}
\label{fig:grainboundary_shape}
\end{center}
\end{figure}

From the optimized Fourier coefficients $a(k)$ and $b(k)$ the energetically favorable grain boundary contour is optained from Eq.~(\ref{eq:perturbation}), which leads to\cite{SMF}
\begin{align} \label{eq:grainboundaryshape}
p(x) &= - \frac{\pi \varepsilon_0 R^2 4 (1+\nu)}{W \beta} \nonumber \\
&\times \frac{\exp \left( - \frac{2 \pi d}{W} \right) \sin \left( \frac{2 \pi (w_0-x)}{W} \right)}{1 - 2 \exp \left( - \frac{2 \pi d}{W} \right) \cos \left( \frac{2 \pi (w_0-x)}{W} \right) +  \exp \left( - \frac{4 \pi d}{W} \right) }.
\end{align}
The result is demonstrated in Fig.~\ref{fig:grainboundary_shape} for different distances between precipitates and grain boundary.
Obviously, a straight interface is favorable for remote precipitates, and grain boundary perturbations become more pronounced for nearby inclusions.

An interesting feature is that the precipitates are not located at symmetry positions of the grain boundary but rather at the left side of the perturbation maxima.
This symmetry breaking emerges from a combination of the shear coupling and the dilatational eigenstrain of the precipitate.
For $\varepsilon_0>0$ the surrounding matrix phase around a precipitate is compressed.
Integration of the shear coupling relation (\ref{eq:shearcoupling}) gives the displacement mismatch $\Delta u_x = \beta p(x)$, which leads to compressive regions according to $\Delta \varepsilon_{xx} = \beta p'(x)$ for regions with positive slope $p'(x)$ and $\beta>0$, in agreement with the optimized precipitate locations in Fig.~\ref{fig:grainboundary_shape}.

The lateral offset $w_0$ appears only in the combination $x-w_0$, reflecting the translational invariance of the problem.

Both the precipitate shape and the elastic free energy have a quadratic dependence on the precipitate radius $R$.
The deformation of the grain boundary therefore increases with the radius, once again constrained by the condition $d \geq R$, such that a crossing of the grain boundary does not occur. 
The initial state of the grain boundary is naturally recovered when the radius vanishes ($R=0$), correctly showing that the grain boundary recovers its shape of a straight line, if no precipitates are present.

\begin{figure}[]
\begin{center}
\includegraphics[width=8.6cm]{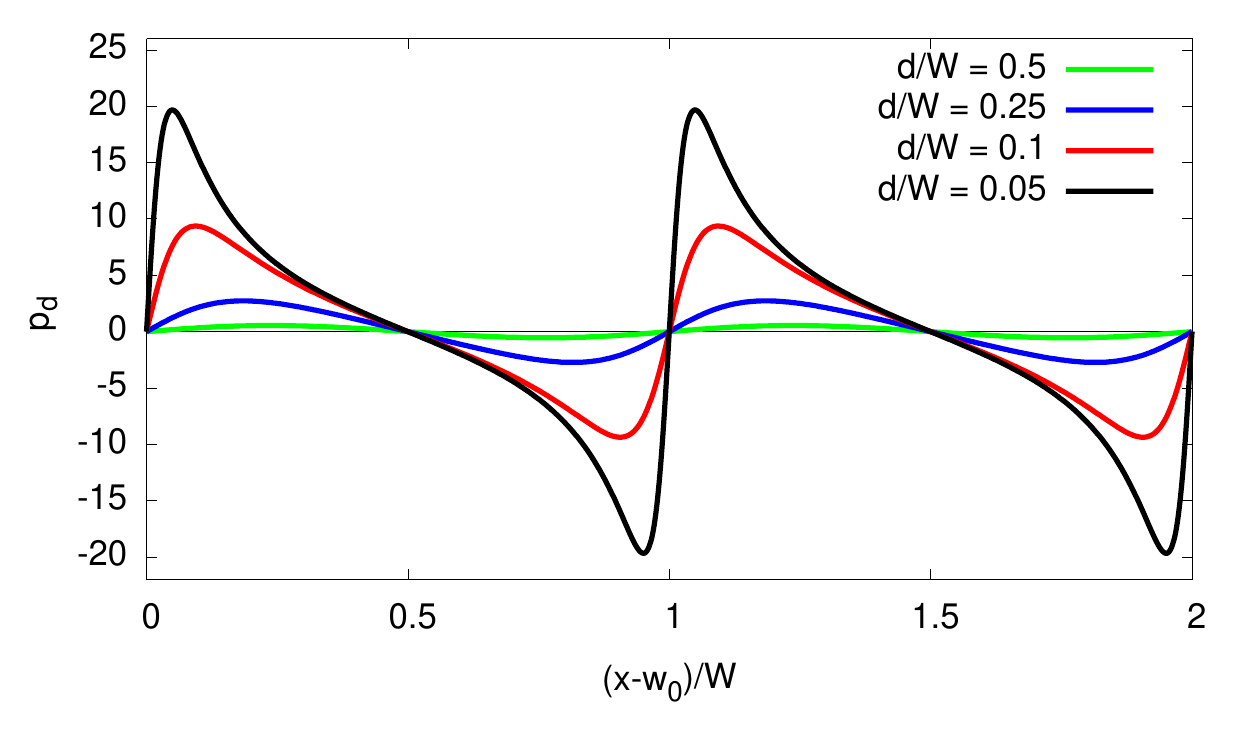}
\caption{The grain boundary shape Eq.~(\ref{eq:grainboundaryshape}) in its dimensionless form, $p_d(x/W) = p(x/W) \cdot (W \beta)/(\varepsilon_0 R^2 (1+\nu)$. 
Different values of $d/W$ are used to show the effect of these two parameters on the grain boundary deformation. 
}
\label{fig:GrainBoundary_Shape_dimless}
\end{center}
\end{figure}

To get a deeper understanding of the functional dependencies of the equilibrium grain boundary contour on the other lengthscales we show it in dimensionless form, $p_d(x/W) = p(x/W) \cdot W \beta/(\varepsilon_0 R^2 (1+\nu))$, in Fig.~\ref{fig:GrainBoundary_Shape_dimless} for different distance ratios $d/W$.
%
It is visible  that the magnitude of the grain boundary deformation increases when $W$ increases.
In extreme cases, when $W$ becomes large, the denominator of Eq.~(\ref{eq:grainboundaryshape}) can become singular, as the exponential functions reach unity.
In this case, the precipitates are far away from each other, such that they can be characterised as independent inclusions.
The result of this extreme case can be interpreted as a break-up of the grain boundary at the locations of the precipitates at $w_0$.
This outcome is in agreement with phase field crystal simulation results\cite{Xu:2016aa}, where it has been demonstrated that a single precipitate can lead to the break-up of a shear-coupled grain boundary.
We note that in between remote precipitates the grain boundary slope scales as $p'(x=w_0 + W/2) \sim 1/W^2$, hence the grain boundary remains essentially flat there.

\section{Solubility limit changes due to shear-coupled motion} \label{sec:solubility}

\begin{figure}
\begin{center}
\includegraphics[width=8.6cm]{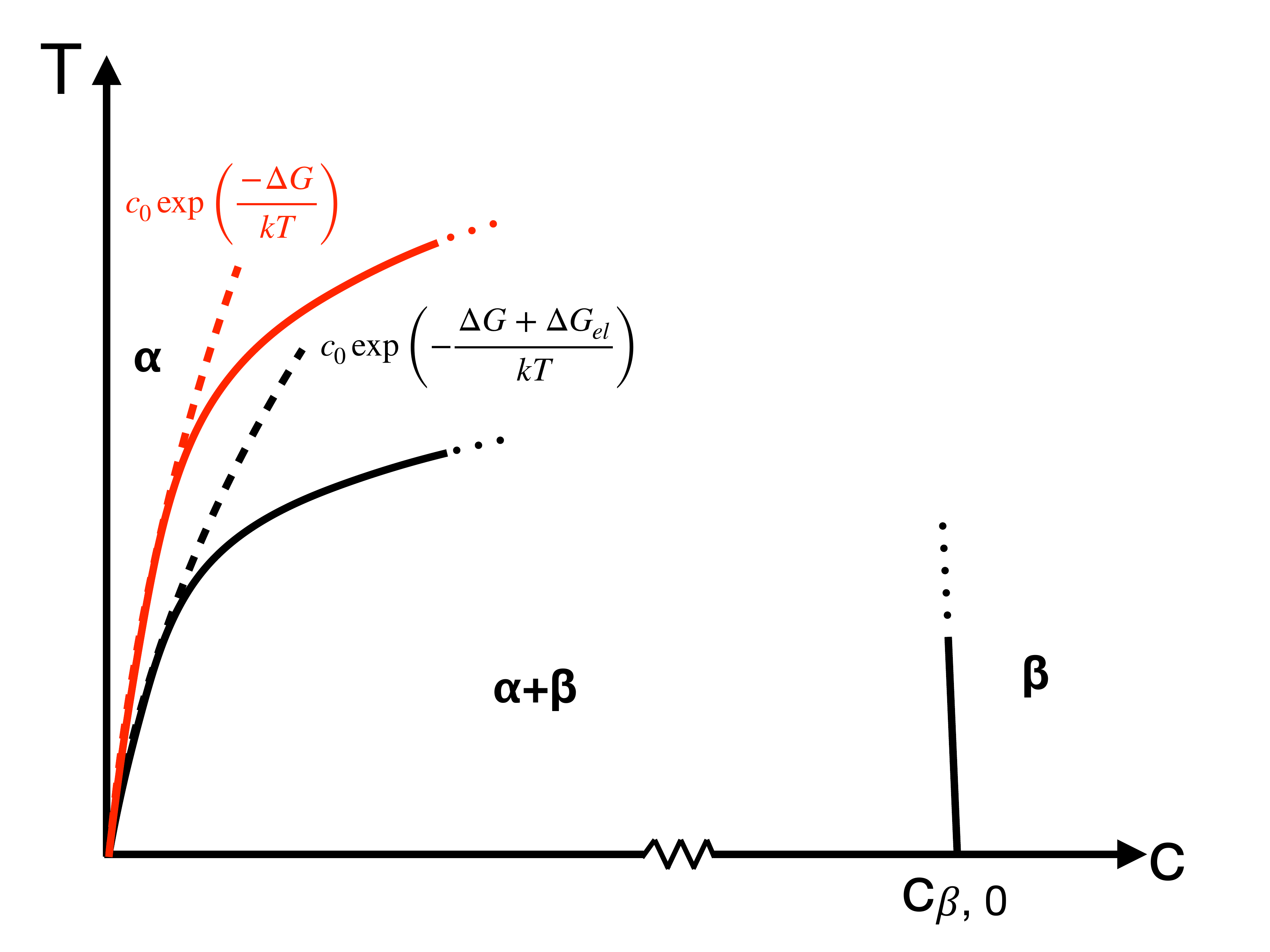}
\caption{Sketch of the phase diagram. In red, the solubility limit without consideration of elastic effects is shown, elastic bulk effects in black. The dashed curves are the Arrhenius approximations according to Eqs. (\ref{exeq1}) and (\ref{exeq2}) for the low temperature limit. The (local) phase diagram near shear coupled grain boundaries lies in between the red and black curves.}
\label{solubility::fig1}
\end{center}
\end{figure}

To understand the influence of shear coupled motion on phase separation, the total Gibbs energy including thermo\-chemical contributions next to the elastic energy has to be considered.
Similar to the analysis in Ref.~\onlinecite{Spatschek:2016aa}, where phase separation with an elastic mismatch in the vicinity of free surfaces has been studied, we focus on binary alloys with coexistence between a disordered solid-solution $\alpha$ phase with zero solubility at $T=0\,\mathrm{K}$ and another $\beta$ phase with coexistence concentration $c_{\beta,0}$ at $T=0$, see Fig.~\ref{solubility::fig1} for a sketch of the phase diagram.
We assume the $\alpha$ phase to be dominated by the configurational entropy contribution (per particle) $g_c\simeq kT c\ln (c/c_0)$ to the Gibbs energy for low solute concentrations $c\ll 1$ and low absolute temperature $T$.
From the asymptotic consideration of the stress free common tangent construction the solubility limit of the $\alpha$ phase is given by an Arrhenius expression ($k$ is the Boltzmann constant)
\begin{equation} \label{exeq1}
c_\alpha(T)=c_0\exp(-\Delta G/kT)
\end{equation}
 with the formation enthalpy difference $\Delta G$, which contains the energetic balance between the phases $\alpha$ and $\beta$, see Ref.~\onlinecite{Spatschek:2016aa} for a detailed discussion.
The inclusion of elastic coherency bulk effects leads --- for the same assumptions as in the present work, i.e.~isotropy, dilatational mismatch and vanishing contrast between the elastic constants between the phases --- to a shift of the formation enthalpy difference according to
\begin{equation} \label{exeq2}
\Delta G \to \Delta G +\Delta G_{el} = \Delta G - \frac{E}{1-\nu} \frac{\varepsilon_0^2}{c_{\beta, 0}} \Omega_\alpha
\end{equation}
with the atomic volume $\Omega_\alpha$  of the pure $\alpha$ phase.
Consequently, the bulk solubility limit of the $\alpha$ phase is therefore increased in comparison to the stress free case.
The central outcome of Ref.~\onlinecite{Spatschek:2016aa} is that near free surfaces elastic stresses can partially relax and therefore attenuate the elastic energy contribution in Eq.~(\ref{exeq2}) by a dimensionless factor $1-\gamma$, i.e.
\begin{equation} \label{exeq3}
\Delta G \to \Delta G + (1-\gamma)  \Delta G_{el}.
\end{equation}
The parameter $\gamma$ has been calculated for a variety of configurations, and for stress relaxation, $0<1-\gamma<1$, the solubility limit of the $\alpha$ phase is therefore decreased in comparison to the bulk coherent phase diagram.
This effect promotes precipitate formation near free surfaces, as expressed through the solubility modification factor\cite{Spatschek:2016aa}
\be \label{eq:solubility}
s = \frac{c_\alpha^\mathrm{surface}(T)}{c_\alpha^\mathrm{bulk}(T)} = \exp\left(\frac{\gamma \Delta G_{el} }{k T}\right).
\ee 
A similar effect can now be expected for phase separation near shear-coupled grain boundaries, which substantiates an earlier effective description\cite{Weikamp:2018aa}.
This morphological degree of freedom allows to reduce the elastic energy when a precipitate forms near the grain boundary, as discussed in the preceding sections.
Here we readily get from Eq.~(\ref{eq:totalenergy_minimised_simplified})
\be \label{eq:gamma}
1 - \gamma = \frac{F^\mathrm{min}}{F^\mathrm{prec}} = 1 - \frac{(1+\nu) \pi^2 R^2}{2 W^2} \frac{1}{\sinh^2\left(\frac{2 \pi d}{W} \right)},
\ee
which is obviously stress relieving, $1-\gamma<1$, and therefore reduces the solubility limit of the $\alpha$ phase against $\beta$ precipitate formation near shear coupled grain boundaries.
The expression (\ref{eq:gamma}) coincides with the dimensionless elastic energy shown in Fig.~\ref{fig:Energy_plot}, which shows that a noticeable reduction of the order of 10\% of the elastic energy is possible.
Remarkably, the result (\ref{eq:gamma}) neither depends on the value of the eigenstrain of the precipitate phase nor on the shear coupling factor, but mainly on the locationss of the precipitates.
Additionally one can observe that the radius still plays a vital role, which is again limited by the constraint $d\geq R$.

In order to estimate the influence of the shear coupling effect on the local solubility limit change we apply the results to the iron-carbon system, which has a phase diagram of the investigated type for the bcc $\alpha$ ferrite--cementite (Fe$_3$C) coexistence for temperatures below about 1000 K.
Whereas the ferrite is a solid solution phase with carbon on octahedral interstitial positions, the cementite appears as stoichiometric phase with a carbon concentration of $c_{\beta, 0}=1/3$ (which equals $6.67$ wt$\%$). 
Using as approximative parameters $T=300$ K, $\varepsilon_0 = 0.0463$, $R = 0.1\,\mu$m, $W = 10\,\mu$m, $E = 175$ GPa, $\nu = 0.25$, $d=R$ and an atomic volume $\Omega_\alpha = 11.78\, \mathrm{\AA}^3$ one obtaines a solubility modification factor of $s \approx 0.5$, hence carbides are expected to precipitate near shear coupled grain boundaries at already at carbon concentration of about half of the bulk solubility limit. 
This effect may have significant implications on the mechanical properties of the steels.

\section{Summary and Conclusions}

Shear coupling of grain boundaries is a mechanism which can lead to mechanical stress relaxation.
As a consequence, attractive interactions between other stress sources like precipitates can result.
We evaluated this effect in the framework of isotropic linear elasticity and coherent, spherical precipitation with a dilatational mismatch to demonstrate the concept.
Small corrugations of the grain boundary provoke (positive) elastic energy due to shear coupling next to the bare, also positive, energy of an array of inclusions.
An energetic cross term, however, can lower the total energy and therefore favor the formation of the precipitates near the grain boundary (attractive interaction).
Energy minimization predicts the strength and range of this short-ranged interaction as well as the corresponding equilibrium grain boundary profile.
A consequence is the local alteration of alloy thermodynamics with a reduced solubility limit near grain boundaries.
Applying these findings to the iron-carbon system allows to qualitatively estimate the effect of the grain boundary and precipitate interaction.
The interstitial model shows, that in the context of shear-coupled motion a correction of the solubility limit of about 50\% at room temperature is possible.  

The serration of grain boundaries in Ni based superalloys\cite{Koul:1983aa,Koul:1985aa,Mitchell:2009aa} is a possible application of the present theory, as similarly argued by Xu et al. \cite{Xu:2016aa}.
The formation of primary $\gamma^{\prime}$ particles during the heat treatment is mandatory for this phenomenon and is a common event in the manufacturing process.
The particles are favored near grain boundaries, while the latter are deformed simultaneously.  
A correlation with primary $\gamma^{\prime}$ size and serration amplitude has been established, which is in agreement with our findings.
The interaction of precipitates and grain boundaries has already been attributed to elastic energy relaxation in Ref.~\onlinecite{Koul:1983aa}, but requires the formation of precipitates at the grain boundary first.
With the present concept of combined precipitation and shear coupled motion one can potentially explain why $\gamma^{\prime}$-particles precipitate near the grain boundary with a simultaneous deformation of the grain boundary contour.


\begin{acknowledgements}
This work has been supported by the DFG via the priority program SPP 1713.
The authors gratefully acknowledge the computing time granted on the supercomputer JURECA at the J\"ulich Supercomputing Centre (JSC).
\end{acknowledgements}


%

\end{document}